# Constraining cosmological dynamics of scalar-tensor models of dark energy in teleparallel gravity

N. Mohammadipour[1]

Department of Physics Education, Farhangian University, P.O. Box 14665-889, Tehran, Iran.

**Abstract**

We consider scalar tensor theory in teleparallel gravity by non-minimally coupling of the generic function of the scalar field $f(\phi)$ to the torsion scalar $T$. Firstly, we obtain the field equations in this regime, then by considering the FRW homogenous cosmology containing a radiation and non- relativistic matter, with energy densities $\rho_r$ and $\rho_m$, we study its evolution as a dynamical autonomous model of dark energy. We show that the cosmological behavior depends on the coupling function $\varsigma = \frac{f'(\phi)}{\sqrt{f(\phi)}}$ and the potential function $\lambda = \frac{V'(\phi)F(\phi)}{V(\phi)f'(\phi)}$. A constant coupling $\zeta$ leads to the quadratic coupling term $f(\phi) \propto (\phi + c)^2$ and $\lambda = $ *constant* gives the power law potential function $V(\phi) \propto f(\phi)^\lambda$ in the scaler field action which is motivated by several mathematical and physical reasons. We study the phase space analysis of these models. We show that $\lambda \gg 1$ is the necessary condition to have radiation era with equation of state $w_{eff} \simeq \frac{1}{3}$ and matter phase with equation of state e$w_{eff} \simeq 0$. Also, the radiation dominated in radiation era and matter dominated in the matter phase can be realized with $\varsigma^2 \ll 1$. Finally, we derive the necessary conditions for the viable cosmological trajectory in this regime.

---

[1] Corresponding email: naser.kurd@cfu.ac.ir

# 1  Introduction

Recently, the f(T) model which is the modification of the old idea of the teleparallel equivalent to General Relativity (TEGR) [1] (This is a theory for the tetrad, whose dynamical equations are equivalent to Einstein equations and instead of using the torsionless Levi-Civita connection one uses the curvatureless Weitzenbock one). Firstly, the f(T) model was proposed to obtain inflation without resorting to an inflation and avoid the Big-Bang singularity [2]. The most of the cosmological implications of this model were studied to explain the accelerated expansion in the universe [3]. Various aspects of these models have been examined in the literature [4]. The violation of local Lorentz invariance and the presence of the new vector degree of freedom in these models have been studied [5]. Teleparallel dark energy has been introduced to by adding a scalar field in the the equivalent, teleparallel, formulation of GR [6]. In this scenario, similar to one type of the scalar tensor gravity, one adds a canonical scalar field, in which the dark energy sector is attributed, allowing also for a non-minimal coupling between the field and the torsion scalar. The observational constraints and phase-space analysis of this model are investigated in [7].

In this paper contrary to teleparallel dark energy, we consider a generic function of scalar field which is coupled to torsion. Using conformal transformation, in addition a minimally coupling between the scalar filed and TEGR, we have a boundary term which violates the local Lorentz invariant and behaves like the more generic theories where an action constructed by a Weitzenbock connection [5].

In the study of the cosmological dynamics of this system, by considering a constant coupling $\varsigma = \frac{f'(\phi)}{\sqrt{f(\phi)}}$, we obtain a quadratic scalar function $f(\phi) \propto (\phi+c)^2$ with the self-interacting power law potential $V(\phi) \propto f(\phi)^\lambda$ ($\lambda$ is a constant parameter). It is worth noting that, these non-minimal coupling function with the effective power law potential is motivated by several mathematical and physical reasons. In particular, the quadratic potential has a strong motivation in inflationary models.

The layout of this paper is as follows. In the following section, we briefly review Teleparallel Equivalent to General Relativity TEGR then, we drive field equations for non-minimally coupling scalar field to TEGR with a kinetic and potential terms in a generic form. Acquire the dynamical equations in a flat FRW universe in order to explain the cosmological applications present in Section 3. We show that, in phase space analysis two function of scalar field appear that by considering this function as a constant, our model constrain by non-minimal coupling term $\zeta\phi^2$ and power law function of self-interacting potential. This form of the coupling term with this potential is motivated by several mathematical and physical reasons [11]-[14]. In Section 4 we investigate the analytical results and critical points of phase space. We show that $\lambda \gg 1$ is the necessary condition to have radiation era with equation of state $w_{eff} \simeq \frac{1}{3}$ and matter phase with equation of state $w_{eff} \simeq 0$. Also, the radiation dominated in radiation era and matter dominated in the matter phase can be realized with $\varsigma^2 \ll 1$. Cosmological applications and the necessary conditions for the viable cosmological trajectory in this regime are concentrated in Section 5. Finally, conclusions are given in section 6.

## 2 Teleparallel gravity and scalar-tensor theories

### 1-2. Teleparallel Equivalent to General Relativity

In this section firstly, we briefly review TEGR in the spatially flat FRW universe. In order to study cosmological dynamics of these theories, we start by the action [8]

$$S = \frac{1}{2k^2} \int e d x^4 T + S_m, \tag{1}$$

where $k^2 = 8\pi \mathcal{G}$, $e = \sqrt{-g} = det(e_\mu^i)$, $S_m$ is the ordinary matter action in the universe and the torsion scalar is the Lagrangian density of teleparallel gravity (TEGR). In this theory, the vierbein field $e_\mu^i$ is the dynamical variable which is related to the metric, $g_{\mu\nu} = \eta_{ij} e_\mu^i e_\nu^i$ 2 In TEGR ones uses the curvatureless Weitzenbock connection $\Gamma^\rho_{\mu\nu} = e_A^\rho \partial_\nu e_\mu^A = -e_\mu^A \partial_\nu e_A^\rho$ and the torsion tensor reads

$$T^\rho_{\mu\nu} = \bar{\Gamma}^\rho_{\nu\mu} - \bar{\Gamma}^\rho_{\mu\nu} = e_A^\rho(\partial_\mu e_\nu^A - \partial_\nu e_\mu^A), \tag{2}$$

where a bar is used to denote all quantities associated with the Weitzenbock connection. The difference between Weitzenbock connection in TEGG and Levi-Civita connection in GR is defined by contorsion $K_\rho^{\mu\nu} = \frac{1}{2}(T_\rho^{\mu\nu} - T_\rho^{\mu\nu} + T_\rho^{\nu\mu})$. The contorsion takes part in the TEGR Lagrangian as

$$T = S_\rho^{\mu\nu} T^\rho_{\mu\nu}, \quad S_\rho^{\mu\nu} = \frac{1}{2}(K_\rho^{\mu\nu} + T_\lambda^{\lambda\mu}\delta_\rho^\nu - T_\lambda^{\lambda\nu}\delta_\rho^\mu) \tag{3}$$

---

[2] here $\mu,\nu$ are the coordinate indices on the manifold while $i,j$ are the coordinate indices for the tangent space of the manifold which all indices run overs 0,1,2,3, also $\eta_{ij}$=diag(1, -1, -1, -1).

The variation with respect to the veribein field of the action Eq.(1) gives equations of motion

$$-4e^{-1}\partial_\mu(eS_A^{\mu\nu}) - 4e_A^\lambda S_\rho^{\mu\nu} T^\rho_{\mu\lambda} + e_A^\nu A = -2k^2 e_A^\rho \hat{T}_\rho^\nu, \tag{4}$$

here) $\hat{T}_\rho^\nu = e_\rho^A \left(e \frac{\partial L_m}{\delta e_\nu^A}\right)$ is the usual energy-momentum tensor. In the spatially flat FRW metric $e_\mu^A$=diag(1, a(t), a(t), a(t)), is an exact solution of Eq.(4), which dose not generate a divergent energy for the whole space-time. In comoving coordinates, the scalar torsion $T$ is $T = -6H^2$ which $H = \frac{\dot{a}}{a}$ is the Hubble parameter. The dynamical equations (4) is equal to equations of motion in *GR*.

## 2-2. Non-minimally coupled scalar-tensor theories in Teleparallel gravity

Consider a non-minimally coupled TEGR scalar field action with canonically normalized scalar field $\phi$ as

$$S = \frac{1}{2k^2} \int dx^4 e[f(\emptyset)T + w(\emptyset)g^{\mu\nu}\partial_\mu\emptyset\partial_\nu\emptyset - 2V(\emptyset)] + S_m(\psi_m; g_{\mu\nu)}.$$

The dynamics of $\phi$ depends on three functions: $f(\phi)$, $\omega(\phi)$ and $V(\phi)$. However, one can always simplify $\omega(\phi)$ and $f(\phi)$ by a redefinition of the scalar field, so that in the Brans-Dicke theory $f(\phi) = \phi$, $\omega(\phi) = \frac{w_{BD}}{\emptyset}$ and in the framework of the teleparallel dark energy, these functions are corresponding to $f(\phi) = 1 + k^2\xi\phi^2$, $\omega(\phi) = k^2$ [6]. By variation of the action (5) we can obtain the field equations in terms of the two functions $f(\phi)$ and $\omega(\phi)$ as the follow, so that any particular choice can be recovered easily:

$$-4[e^{-1}\partial_\mu(eS_A^{\mu\nu}) + e_A^\lambda S_\rho^{\mu\nu} T^\rho_{\mu\lambda} - \frac{1}{4}e_A^\nu T]f(\emptyset) - 4S_A^{\mu\nu}\partial_\mu\emptyset f'(\emptyset) +$$
$$e_A^\nu(w(\emptyset)g^{\mu\lambda}\partial_\lambda\phi\partial_\nu\emptyset - 2V(\emptyset)) - w(\emptyset)e_A^\mu\partial_\mu\emptyset\partial^\nu\emptyset = -2k^2 e_A^\rho \hat{T}_\rho^\nu, \tag{6}$$

$$F'(\phi)T - \omega'(\phi)(\partial_\alpha\phi)^2 - 2\omega(\phi)\bar{\square}\phi - 2V'(\phi) = 0, \tag{7}$$

where a prime denotes derivative with respect to $\phi$ and $\bar{\square}\phi = \partial_\mu + \bar{\Gamma}^\rho_{\rho\mu}$. It can be easily derived the conservation equation as

$$\nabla_\mu \hat{T}^\mu_\nu = 0. \tag{8}$$

The above equations are written in the so-called Jordan frame. It is worth noting that, the Jordan metric $g_{\mu\nu}$ which is coupled to matter in action Eq.(5) defines the times and the lengths measured by laboratory clocks and rods. In particular, the redshift and the observed Hubble parameter are quantities in this frame.

The f(T) model with general action

$$S = \frac{1}{2k^2} \int e \, dx^4 f(T), \tag{9}$$

firstly, was proposed to obtain inflation without inflaton and avoid the Big Bang singularity [2], then the cosmological applications studied in the more literatures [3]-[5]. One can use the auxiliary fields A and B and rewrite the action Eq.(9) as the follow

$$S = \frac{1}{2k^2} \int e \, dx^4 [(T-A)]B + f(A)], \tag{10}$$

Varying with respect to A and B give B = f'(A) and A = T, respectively.

Choosing $f(\phi) = f'(A)$, the action Eq.(5) covers f(T) gravity with $\omega(\phi) = 0$

And $$V(\phi) = \frac{\phi f'(\phi) - f(\phi)}{2}$$

With a suitable conformal transformation on metric, we can change the non-minimally scalar-tensor (NST) Lagrangian into the Einstein-Hilbert one, minimally coupled to a scalar field. It is worth noting that, any NST theory is *mathematically* equivalent to General Relativity with the scalar field. Like what was said, to change the NST Lagrangian teleparallel gravity into TEGR with the scalar field, we have a scalar-torsion coupling term [9]. However, this boundary term implies that actions of the form given in Eq.(5) when change from the original frame so-called Jordan frame into TEGR one, minimally coupled to a scalar field, are not locally Lorentz invariant. So, they are not special as the TEGR, but behave like the more generic theories where an action constructed by a Weitzenbock connection. In the next section we shall study the cosmological dynamics for the JF frame action Eq.(5) in the presence of a perfect fluid.

## 3 Homogeneous cosmology

We assume the background to be a perfect fluid with radiation and dust matter, whose energy densities $\rho_r$ and $\rho_m$ which satisfy the usual continuity equations. By imposing the FRW geometry $ds^2 = dt^2 - a^2(t)dx^2$, where $t$ is cosmic time and $a(t)$ is the scale factor, from equations (6)-(8), we obtain the Friedmann, continuity and field equations as the follow

$$3f(\emptyset)H^2 = \rho_m + \rho_r + \frac{1}{2}\dot{\emptyset}^2 + V(\emptyset) \tag{11}$$

$$-2Hf(\emptyset) = \rho_m + \frac{3}{4}\dot{\rho_r} + \dot{\emptyset}^2 + 2H\dot{\emptyset}f'(\emptyset) \tag{12}$$

$$(\ddot{\emptyset} + 3H\dot{\emptyset} + 3H^2 f'(\emptyset) + V'(\emptyset) = 0 \tag{13}$$

$$\dot{\rho_m} = 3H\rho_m = 0 \tag{14}$$

$$\dot{\rho_r} + 4H\rho_r = 0, \tag{15}$$

here, $k^2 = 8\pi G = 1$ and $\omega = 1$. Eqs.(11-15) determine the dynamics of the system Eq.(5) in a homogeneous and isotropic background. These equations correspond to the most general parametrization Eq.(5) of scalar-tensor theories, many cases are easily recovered. For example, the Teleparallel dark energy [6] is obtained for $f(\phi) = 1 + \xi\phi^2$ and $\omega = 1$.

### 1-3. Cosmological dynamics

In order to confront the dark energy (DE) equation of state with observation, we can rewrite Eqs.(11), (12) in a more convenient form as [9]:

$$3f_0 H^2 = \rho_m + \rho_r + \rho_\emptyset \tag{16}$$

$$-2\dot{H}f_0 = \rho_m + \frac{4}{3}\rho_r + \rho_\emptyset + P_\emptyset, \tag{17}$$

where, we can naturally define gravitationally induced forms of dark energy density and pressure as

$$\rho_\emptyset = \frac{1}{2}\dot{\emptyset}^2 + V(\emptyset) - 3H^2(f - f_0) \tag{18}$$

$$P_\emptyset = \dot{\emptyset}^2 + 2H\dot{\emptyset}f'(\emptyset) + 2\dot{H}(f - f_0) - \rho_\emptyset, \tag{19}$$

and the subscript (0) represents present day values. In this way, the continuity equation of dark energy holds

$$\dot{\rho}_\emptyset + 3H(\rho_\emptyset + P_\emptyset) = 0, \tag{20}$$

with the equation of state which is defined as

$$w_\emptyset = \frac{P_\emptyset}{\rho_\emptyset} = -1 + \frac{\dot{\emptyset}^2 + 2H\dot{\emptyset}f'(\emptyset) + 2\dot{H}(f-f_0)}{\frac{1}{2}\dot{\emptyset}^2 + V(\emptyset) - 3H^2(f-f_0)}. \tag{21}$$

Let us define the following dimensionless variables

$$x_1^2 = \frac{\rho_r}{3fH^2}, \quad x_2^2 = \frac{\dot{\emptyset}^2}{6fH^2}, \quad x_3^2 = \frac{V}{3fH^2} \tag{22}$$

and

$$\Omega_r = x_1^2, \qquad \Omega_\emptyset = x_2^2 + x_3^2, \qquad \Omega_m = 1 - x_1^2 - x_2^2 - x_3^2. \tag{23}$$

Then Eqs.(11-15) can be rewrite as a first order dynamical system

$$\frac{dx_1}{dN} = -x_1\left(2 + \frac{\dot{H}}{H^2} + \frac{\sqrt{6}}{2}\varsigma x_2\right), \tag{24}$$

$$\frac{dx_2}{dN} = -x_2\left(3 + \frac{\dot{H}}{H^2} + \frac{\sqrt{6}}{2}\varsigma x_2\right) - \frac{\sqrt{6}}{2}\varsigma(1 + \lambda x_3^2) \tag{25}$$

$$\frac{dx_3}{dN} = x_3\left(\frac{\sqrt{6}}{2}\varsigma(\lambda - 1)x_2 - \frac{\dot{H}}{H^2}\right) \tag{26}$$

where $N = \ln\left(\frac{a}{a_i}\right)$ ($a_i$ is the initial value of the scale factor), $\varsigma = \frac{f'(\emptyset)}{\sqrt{f(\emptyset)}}$, $\lambda = \frac{V'(\emptyset)F(\emptyset)}{V(\emptyset)f'(\emptyset)}$ and

$$\frac{\dot{H}}{H^2} = -\frac{3}{2} - \frac{1}{2}x_1^2 - \frac{3}{2}x_2^2 - \sqrt{6}\varsigma x_2 + \frac{3}{2}x_3^2. \tag{27}$$

From Eq.(24), (26) which indicate that $x_1 = 0$ and $x_3 = 0$ are invariant submanifolds of the dynamical system. These show that there is no orbit for which the radiation of fluid or the potential of scalar field can become exactly zero. This means that, if the radiation or potential is initially set to zero, it will remain zero.

We can rewrite the effective equation of state $w_{eff} = -1 - \frac{2\dot{H}}{3H^2}$ and the deceleration parameter $q = -1 - \frac{\dot{H}}{H^2}$ in terms of the dynamical variables Eq.(22) as:

$$w_{eff} = x_2^2 + \frac{2\sqrt{6}}{3}\varsigma x_2 + \frac{1}{3}x_1^2 - x_3^2, \tag{28}$$

$$q = \frac{1}{2}(1 + 3x_2^2 + 2\sqrt{6}\varsigma x_2 + x_1^2 - 3x_3^2). \tag{29}$$

Considering $\zeta$ and $\lambda$ as constants, one can obtain

$$f(\emptyset) = \frac{\zeta^2}{4}(\emptyset + c)^2, \quad V(\emptyset) = nf(\emptyset)^\lambda, \tag{30}$$

where, n is a constant. The cosmological dynamics and the phase plane of the flat FRW cosmological model with the coupling term $\zeta\emptyset^2 R$ and the potential function $V(\phi) \propto \emptyset^n$ have been investigated in [11, 12]. Furthermore, the minimally coupled scalar field with potential $V(\phi) \propto \emptyset^2$ has a strong motivation in inflationary models. Its Generalizations with non-minimal coupling term $\zeta\emptyset^2 R$ with quadratic potential function [13] and $(k^{-2} - \zeta\emptyset^2)$ with a self-interacting potential $V(\phi) \propto \emptyset^4$ [14] have been studied in the context of the origin of the canonical inflaton field. However, the phase space analysis of teleparallel dark energy which is non-minimally coupled scalar field with the coupling term $(\zeta\emptyset^2 + 1)T$ with an exponential self-interacting potential have been investigated in [15]. It is worth noting that, for a generic function of $f(\Phi)$ which is non-minimal coupled to $T$, considering $(\frac{\zeta}{2})^2\Phi^2 T$ with $\lambda$ = *constant* leads to a potential function $V(\Phi) \propto \Phi^{2\lambda}$ Eq.(30) with redefinition $\Phi = \phi + c$.

Now, we study the cosmological dynamics of non-minimally coupled scalar-tensor theory in teleparallel gravity that is reconstructed by $\zeta$ and $\lambda$ which are constant.

## 4  Analytical results and critical points

The field equations (24-26) which are written in terms of the dynamical variable Eq.(22) makes it easy to look at the analytical properties of them. The study of phase space is the investigation of the evolution of the cosmologically trajectories by finding the critical points and the stability conditions of them. The fixed points are where the r.h.s. of the Eqs. (24-26) is zero. We will assume two cases: i) we have a non-relativistic dust like matter in the absence of radiation ($\Omega_r = x_1^2 = 0$) ii) we have radiation and dust matter.

### 1-4. Dust matter in the absence of radiation

From Eq.(24) $x_1 = 0$ (in the absence of radiation) is an invariant sub_manifold. This tells us that the system can not go through the subspace $x_1 = 0$ and can only approach it asymptotically. The critical points with cosmological quantities for this case are summarized in Table1. Firstly, we shall discuss the case $\zeta = 0$, which corresponds to a standard minimally coupled scalar field.

Choosing $\zeta = 0$, the critical point $E_1$ corresponds to the dust matter dominated point ($x_2 = 0$, $x_3 = 0$, $w_{eff} = 0$) with the eigenvalues ($\frac{3}{2}, -\frac{3}{2}$), which show that this point is a saddle point. C1 and D1 correspond to ($x_2 = \pm 1$, $x_3 = 0$, ωeff = 1) with eigenvalues (3, −3). These are saddle points, that correspond to solutions where the kinetic energy of the scalar field is dominated in Eq.(11). Also, the point $B_1$ indicates ($x_2 = 0$, $x_3 = \pm 1$, $w_{eff} = -1$) with eigenvalues (−3, −3), which is a stable point. The viable cosmological trajectory for $\zeta = 0$ and constant $\lambda$ corresponds to.

Table 1: The critical points and the cosmological quantities of them for dynamical system Eqs.(24-26) in the absence of radiation ($x_1 = 0$). sequence from the matter dominated point $E_1$ to the scalar-field dominated $B_1$.

| point | $x_2$ | $x_3$ | $\frac{\dot{H}}{H^2}$ | $w_{eff}$ | $q$ | $\Omega_\Phi$ | $\Omega_m$ |
|---|---|---|---|---|---|---|---|
| $A_1$ | $-\frac{\sqrt{6}}{2\zeta\lambda}$ | $\pm\frac{\sqrt{6-4\zeta^2\lambda}}{2\zeta\lambda}$ | $-\frac{3(\lambda-1)}{2\lambda}$ | $-1+\frac{\lambda-1}{\lambda}$ | $-1+\frac{3(\lambda-1)}{2\lambda}$ | $\frac{3-\zeta^2\lambda}{\zeta^2\lambda^2}$ | $1-\frac{3-\zeta^2\lambda}{\zeta^2\lambda^2}$ |
| $B_1$ | $-\frac{\zeta(\lambda+1)}{\sqrt{6}}$ | $\pm[1-\frac{\zeta^2(1+\lambda)^2}{6}]^{\frac{1}{3}}$ | $\frac{\zeta^2(1-\lambda^2)}{2}$ | $-[1+\frac{\zeta^2(1-\lambda^2)}{2}]$ | $-[1+\frac{\zeta^2(1-\lambda^2)}{2}]$ | 1 | 0 |
| $C_1$ | 1 | 0 | $-(3+\sqrt{6}\zeta)$ | $1+\frac{2\sqrt{6}\zeta}{3}$ | $2+\sqrt{6}\zeta$ | 1 | 0 |
| $D_1$ | -1 | 0 | $-3+\sqrt{6}\zeta$ | $1-\frac{2\sqrt{6}\zeta}{3}$ | $2-\sqrt{6}\zeta$ | 1 | 0 |
| $E_1$ | $-\frac{\sqrt{6}\zeta}{3}$ | 0 | $-\frac{3}{2}+\zeta^2$ | $-\frac{2}{3}\zeta^2$ | $\frac{1}{2}-\zeta^2$ | $\frac{2}{3}\zeta^2$ | $1-\frac{2}{3}\zeta^2$ |

The cosmological quantities, stability and existence conditions of them are summarized in Table2.

These points are obtained from the phase space investigation of the standard quintessence with exponential potential [16]. It is worth noting that, in this case we do not have any scaling solution. We have a saddle matter dominated point, two saddle points correspond to kinetic dominated and the attractor scalar field dominated solution, that they are independent of any values of $\lambda$ and $\zeta$.

Now, we consider the case of non-zero values of $\zeta$. In this case we have five fixed points.

$A_1$: *scaling solution with matter*√#
$(x_2 = -\frac{\sqrt{6}}{2\zeta\lambda}, x_3 = \pm\frac{\sqrt{6-4\zeta^2\lambda}}{2\zeta\lambda}, w_{eff} = -\frac{3(\lambda-1)}{2\lambda})$

This point is the scaling solution with matter, which exists for $\zeta^2 \lambda \leq \frac{3}{2}$. Note that, the matter era ($w_{eff} \simeq 0$) can be realized by this point with $\lambda \gg 1$.

$B_1$: *scalar field-dominated*

| point | $x_2$ | $x_3$ | $\frac{\dot{H}}{H^2}$ | $\omega_{eff}$ | $q$ | $\Omega_\Phi$ | $\Omega_m$ | Stability | Existence |
|---|---|---|---|---|---|---|---|---|---|
| $A_1$ | 0 | 0 | $-\frac{3}{2}$ | 0 | $\frac{1}{2}$ | 0 | 1 | Saddle point | All $\lambda$ |
| $B_1$ | 0 | 1 | 0 | $-1$ | $-1$ | 1 | 0 | Stable | All $\lambda$ |
| $C_1$ | ±1 | 0 | $-3$ | 1 | 2 | 1 | 0 | Saddle point | All $\lambda$ |

Table 2: The critical points, the cosmological quantities, Stability and the existence conditions of dynamical system Eqs.(24-26) with $x_1 = \zeta = 0$.

$$(x_2 = -\frac{\zeta(\lambda+1)}{\sqrt{6}}, x_3 = \pm\sqrt{1 - \frac{\zeta^2(1+\lambda)^2}{6}}, w_{eff} = -1 - 3\zeta^2(1-\lambda^2))$$

$B_1$ is the scalar field dominated solution. The existence condition of these solutions is $\zeta^2 \leq \frac{6}{(1+\lambda)^2}$. For $\lambda = 1$, the point $B_1$ reduce to De sittr point ( $x_2 = -\frac{2\zeta}{\sqrt{6}}$, $x_3 = \pm\sqrt{1 - \frac{2\zeta^2}{3}}$, $w_{eff} = -1$ )with the eigenvalues

$$a_1 = -3 + \zeta^2, \qquad a_2 = -3 + 2\zeta^2. \qquad (31)$$

It is straightforward that, the attractor condition is $\zeta^2 \leq \frac{3}{2}$ and when $\zeta > 3$ we have unstable point. It is worth noting that, the self-interaction potential is taken quadratic form $V(\Phi) \propto \Phi^2$, that has a strong motivation in inflationary models. Also, $\lambda = -1$ gives another de Sitter point ($x_2 = 0, x_3 = \pm 1, w_{eff} = -1$) with the eigenvalues

$$a_1 = -3 + \sqrt{6}\zeta, \qquad a_2 = -3. \qquad (32)$$

In this case, we have an attractor point when $\zeta < \frac{\sqrt{6}}{2}$ otherwise, $B_1$ is a saddle point. The potential function leads to $V(\Phi) \propto \Phi^{-2}$. The stability and the existence conditions of this point are summarized in Table3.

$C_1$ : *Kinetic-dominated*

$(x_2 = 1, x_3 = 0, w_{eff} = 1 + \frac{2+\sqrt{6}\zeta}{3})$

This point is one of the Kinetic-dominated solution, that the eigenvalues of the linearized system are:

$3+\sqrt{6}\zeta, 3+\frac{\sqrt{6}\zeta}{2}(\lambda + 1)$

| point | Stability conditions | Existence |
|---|---|---|
| $C_1$ | Unstable for $\zeta > 0, \lambda > -\left(1 + \frac{\sqrt{6}}{\zeta}\right)$ and $-\frac{\sqrt{6}}{2} < \zeta < 0, \lambda < -(1 + \frac{\sqrt{6}}{\zeta}) <$ | |
| | Saddle point for $\zeta < -\frac{\sqrt{6}}{2}, \lambda < -(1 + \frac{\sqrt{6}}{\zeta})$ | |
| | Stable for $\zeta < -\frac{\sqrt{6}}{2}, \lambda < -(1 + \frac{\sqrt{6}}{\zeta})$ | All $\lambda, \zeta$ |
| $D_1$ | Unstable for $\zeta > -\frac{\sqrt{6}}{2}, \lambda < -(1 - \frac{\sqrt{6}}{\zeta})$ | |
| | Saddle point for $\zeta > 0, \lambda > -\left(1 + \frac{\sqrt{6}}{\zeta}\right)$ and $-\frac{\sqrt{6}}{2} < \zeta < 0, \lambda < -(1 + \frac{\sqrt{6}}{\zeta})$ | All $\lambda, \zeta$ |
| | Stable for $\zeta < 0, \lambda > -\left(1 + \frac{\sqrt{6}}{\zeta}\right)$ and | |
| $E_1$ | Unstable for $\zeta^2 > \frac{3}{2}, \lambda < \frac{2}{2\zeta^2}$ | All $\lambda, \zeta$ |
| | Saddle point for $0 < \zeta^2 < \frac{3}{2}, \lambda > \frac{3}{2\zeta^2}$ | |
| | Attractor for $\zeta^2 < 0, \lambda > \frac{3}{2\zeta^2}$ | |

Table 3: The matrix perturbation eigenvalues $\alpha_i$ and the stability and exis_ tence conditions of the critical points Table1 which are obtained from ana_ lytical results of phase space studying of the autonomous system Eqs.(24)-(26).

$D_1$: *Kinetic-dominated*   √

(

$x_2 = -1, x_3 = 0, w_{eff} = 1 - \frac{2\sqrt{6}\zeta}{3})$

$D_1$ is another Kinetic-dominated solution with the eigenvalues of the

linearized system

$$-3+\sqrt{6}\zeta, \quad 3 - \frac{\sqrt{6}\zeta}{2}(\lambda + 1)$$

- $E_1$ : $\phi$ matter-dominated ($\phi$MDE )
  $(x_2 = -\frac{\sqrt{6}\zeta}{3}, x_3 = 0, w_{eff} = \frac{2\zeta^2}{3}$

  This point is the $\phi$ matter-dominated era [17, 18], that the eigenvalues #
  of the linearized system are:

$$-\frac{3}{2} + \zeta^2, \frac{3}{2} - \zeta^2 \lambda$$

The stability and existence conditions of the critical points $C_1$, $D_1$ and $E_1$ are summarized in Table3.

### 2-4. Dust matter with radiation

| point | $x_1$ | $x_2$ | $x_3$ | $w_{eff}$ | $\Omega_\Phi$ | $\Omega_r$ | $\Omega_m$ |
|---|---|---|---|---|---|---|---|
| $R_1$ | $\sqrt{\frac{\lambda+1}{\lambda} - \frac{4}{\zeta^2\lambda^2}}$ | $-\frac{2\sqrt{6}}{3\zeta\lambda}$ | $\sqrt{\frac{4}{3\zeta^2\lambda^2} - \frac{1}{\lambda}}$ | $-1 + \frac{4(\lambda-1)}{3\lambda}$ | $\frac{4}{\zeta^2\lambda^2} - \frac{1}{\lambda}$ | $\frac{\lambda+1}{\lambda} - \frac{4}{\zeta^2\lambda^2}$ | 0 |
| $R_2$ | $\sqrt{1 - \frac{3\zeta^2}{2}}$ | $-\frac{\sqrt{6}\zeta}{2}$ | 0 | $\frac{1}{3} - \zeta^2$ | $\frac{3\zeta^2}{2}$ | $1 - \frac{3\zeta^2}{2}$ | 0 |

Table 4: The critical points and the cosmological quantities of them for dynamical system Eqs.(24-26) with $x_1 \neq 0$.

In this subsection we have studied dynamics of the system Eq.(24-26) with radiation and non-relativistic dust like matter. In this case, we have the fixed points which are presented in previous subsection. Also, we obtain a new critical point that, it is summarized with properties in Table4. Now, we investigate the matrix perturbation eigenvalues and the stability conditions for this point.

$R_1$: *Scaling solution with radiation*√#

$$\left(x_1 = \sqrt{\frac{\lambda+1}{\lambda} - \frac{4}{\zeta^2\lambda^2}}, x_2 = -\frac{2\sqrt{6}}{3\zeta\lambda}, x_3 = \sqrt{\frac{4}{3\zeta^2\lambda^2} - \frac{1}{\lambda}}, w_{eff} = -1 + \frac{4(\lambda-1)}{3\lambda}\right),$$

$R_1$ is the scaling solution with radiation. This point exists when $\frac{4}{\lambda(\lambda+1)} \leq \zeta^4 \leq \frac{2}{3\lambda}$. It is worth noting that, if this point is responsible for the radiation phase, the condition $\lambda \gg 1$ is required which also is the matter era condition in scaling solution $A_1$ with equation of state, $w_{eff} \simeq 0$.

$R_2 : \phi$ radiation dominated point

$$\left(x_1 = \sqrt{1 - \frac{3\zeta^2}{2}}, x_2 = -\frac{\sqrt{6}\zeta}{2}, x_3 = 0, w_{eff} = \frac{1}{3} - \zeta^2\right)$$

$R_2$ is the $\phi$ radiation dominated era. The eigenvalues of the perturbation matrix are

$$a_1 = 2 - \frac{3\zeta^2\lambda}{2}, a_i = -\frac{1}{4}[\sqrt{6}\zeta \pm \sqrt{2\zeta(3\zeta + 4\sqrt{6}) - 12\zeta^3(3\zeta + \sqrt{6})}], \quad (33)$$

#where $i = 2,3$. Considering $\zeta^2 \ll 1$ this solution can lead to the radiation dominated era with equation of state, $w_{eff} \simeq \frac{1}{3}$ and $\Omega_r \simeq 1$. In this case, the eigenvalue $a_1 > 0$ and we have unstable or saddle point.

It is worth noting that, in phase space analysis of action Eq.(5) we obtain a $\phi$ radiation dominated point $R_2$ ($\phi$RDE), a $\phi$ matter dominated point $E_1$ ($\phi$MDE), a scaling solution with matter $A_1$, a scalar-field dominated solution $B_1$ with two kinetic-dominated points $C_1$ and $D_1$. We note that, $x_3 = 0$ in the point $R_2$ gives invariant submanifold of this dynamical system. Since, when system leaves the radiation-dominated era and fall on to the matter phase, $x_3$ will remain zero.

.

# 5 Cosmological consequences

We perform a thorough phase space analysis on the scalar-tensor models of dark energy in teleparallel gravity. In order to have a matter era which followed by the late-time attractor firstly, assume that we have non-relativistic dust like matter. The fixed points and their quantities which are extracted with this assumption are summarized in Table1. For the minimally coupling case, these points reduce to four points that are summarized in Table2. $A_1$ is a saddle matter dominated point which is followed by accelerated stable point $B_1$. These points with two kinetic-dominated points are derived in standard quintessence [16] and teleparallel dark energy [15]. In the non-minimal case, we obtain a scaling solution, a scalar field dominated point, two points which represent kinetic-dominated solutions and a $\phi$ matter dominated point similar to standard scalar-tensor models of dark energy [18]. In the absence of radiation, the cosmological trajectory for constant $\lambda$ corresponds to the sequence from the point $E_1$ to the scalar- field dominated point $B_1$ under the condition $\zeta^2 \ll 1$ and $\lambda > \frac{3}{2\zeta^2}$. However, considering radiation with non-relativistic dust like matter, the cosmological trajectory start in radiation dominated point $R_2$ then go through the matter dominated point $E_1$ which is followed by attractor point $B_1$ under the conditions $\zeta^2 \ll 1$ and $\lambda > \frac{3}{2\zeta^2}$. We note that, if $\lambda^2 \gg 1$ the scaling solutions $A_1$ and $R_1$ represent the radiation and matter eras with $w_{eff} \simeq \frac{1}{3}$ and $w_{eff} \simeq 0$, respectively.

# 6 Conclusions

In this paper we have derived conditions for cosmologically viable scalar-tensor models of dark energy in teleparallel gravity. Considering a generic form of scalar-tensor action Eq.(5), we obtained the field equations. In the spatially flat FRW metric with this assumption that the background to be a perfect fluid with radiation and dust matter, with energy densities $\rho_r$ and $\rho_m$, the Friedmann, continuity and field equations are obtained. By recasting the field equations into dimensional autonomous system in to cases i) in the absence of radiation ii) non-relativistic dust like matter with radiation, we investigated the cosmological dynamics of them. Performing phase space analysis of these models for constants $\lambda$ and $\zeta$ which are $\varsigma = \frac{f\prime(\phi)}{\sqrt{f(\phi)}}$ and $\lambda =$ $\frac{V'(\phi)F(\phi)}{V(\phi)f'(\phi)}$ led to the quadratic coupling term $f(\phi) \propto (\phi + c)^2$ and the power law potential function $V(\phi) \propto f(\phi)^\lambda$. For $\zeta = 0$ this model covers the standard quintessence solutions [16]. It is worth noting that, this model has a mathematically and physically interesting behavior in scaling solution with radiation and matter in early-time and late time. So that, from scaling solution $A_1$ we have an attractor scaling solution with $w_{eff} = -1$ for $\lambda = 1$. But, the cosmological trajectory start in radiation dominated point $R_2$ then go through the matter dominated point $E_1$ which is followed by attractor point $B_1$ under the conditions $\zeta^2 \ll 1$ and $\lambda > \frac{3}{2\zeta^2}$.

# References

[1] A. Einstein 1928, Sitz. Preuss. Akad. Wiss. p. 217; ibid p. 224; A. Unzicker and T. Case,

[2] R. Ferraro and F. Fiorini, Phys. Rev. D 75, 084031 (2007).

[3] G. R. Bengochea and R. Ferraro, Phys. Rev. D 79, 124019 (2009);

[4] P. Wu and H. W. Yu, Phys. Lett. B 693, 415 (2010);

[5] B. Li, T. P. Sotiriou and J. D. Barrow, Phys. Rev. D 83, 064035 (2011); M. Li, R. X. Miao and Y. G. Miao, JHEP 1107, 108 (2011).

[6] C. Q. Geng, C. C. Lee, E. N. Saridakis, Y. P. Wu, Phys. Lett. B 704, 384 (2011).

[7] C. Q. Geng, C. C. Lee and E. N. Saridakis, arXiv:1110.0913;

[8] V. C. de Anderade, L. C. T. Guillen and J. G. Periera, arXiv:0011087v1.

[9] S. Tsujikawa, P. Rev. D 76, 023514 (2007).

[10] R. J. Yang, Europhys. Lett. 93, 60001 (2011).

[11] S. Carloni, J. A. Leach, S. Capozziello and P. K. S. Dunsby, Class. Quant. Grav 25, 035008 (2008).

[12] O. Hrycyna and M. Szydlowski, JCAP 04, 026 (2009).

[13] S. C. Park and S. Yamaguchi, JCAP 0808, 009 (2008).

[14] H. Noh and J. Hwang, arXiv:0102311[astro-ph].

[15] C. Xu, E. N. Saridakis and G. Leon, arXiv:1202.3781.

[16] E. j. Copeland, A. R. Liddle and D. Wands, Phys. Rev. D 57, 4686 (1998).

[17] L. Amendola, Phys. Rev. D 62, 043511 (2000).

[18] S. Tsujikawa, K. Uddin, S. Mizuno, R. Tavakol and J. Yokoyama, Phys. Rev. D 77, 103009 (2008).